\documentclass[twocolumn,doublespace,showkeys,showpacs,superscriptaddress]{IEEEtran}

\usepackage{graphicx,epic,eepic,epsfig,amsmath,latexsym,amssymb,verbatim,subfigure,color}
\usepackage{theorem}

\newcommand{\ket}[1]{|{#1}\rangle}
\newcommand{\bra}[1]{\langle{#1}|}
\newcommand{\X}{\sigma_x}
\newcommand{\Z}{\sigma_z}
\newcommand{\Y}{\sigma_y}

\newcommand{\iden}{1 \hspace{-1.0mm}  {\bf l}}

\newcommand{\ncd}{\newcommand}
\ncd{\QC}{$\mbox{QC}_{\cal{C}}\;$}
\ncd{\QCpr}{${\mbox{QC}_{\cal{C}}}^\prime\;$}
\ncd{\QCns}{$\mbox{QC}_{\cal{C}}$}
\ncd{\QCprns}{${\mbox{QC}_{\cal{C}}}^\prime$}
\ncd{\cskN}{{|\phi_{\{\kappa\} } \rangle}_{{\cal{C}}_N}}
\ncd{\cskNpr}{{|\phi_{\{\kappa^\prime\} } \rangle}_{{\cal{C}}_N}}
\ncd{\cskNtil}{{|\phi_{\{\tilde{\kappa} \} } \rangle}_{{\cal{C}}_N}}
\ncd{\csk}{{|\phi_{\{\kappa\} } \rangle}_{\cal{C}}}
\ncd{\csktil}{{|\phi_{\{\tilde{\kappa} \} } \rangle}_{\cal{C}}}
\ncd{\cskf}{|\phi_{\{\kappa\} } \rangle_{\cal{C}}}
\ncd{\csktilf}{|\phi_{\{\tilde{\kappa} \} } \rangle_{\cal{C}}}
\ncd{\bracsk}{\mbox{}_{\cal{C}}\langle\phi_{\{\kappa\} }|}
\ncd{\bracsktil}{\mbox{}_{\cal{C}}\langle\phi_{\{\tilde{\kappa} \} }|}
\ncd{\nbracsk}{\mbox{}_{\cal{C}}\langle\phi_{\{\kappa\} }}
\ncd{\nbracsktil}{\mbox{}_{\cal{C}}\langle\phi_{\{\tilde{\kappa} \} }}
\ncd{\cs}{|\phi \rangle_{\cal{C}}\;}
\ncd{\csns}{|\phi \rangle_{\cal{C}}}
\ncd{\nbgh}{\text{nbgh}}
\ncd{\Sab}{S^{ab}}
\ncd{\Sba}{S^{ba}}
\ncd{\ds}{\displaystyle}
\ncd{\ovl}{\overline}

\newtheorem{fact}{Fact}
\newtheorem{definition}{Definition}
\newtheorem{example}{Example}

\newtheorem{theorem}{Theorem}
\newtheorem{corollary}{Corollary}

\newcommand{\nc}{\newcommand}
\nc{\rnc}{\renewcommand}
\nc{\beq}{\begin{equation}}
\nc{\eeq}{{\end{equation}}}
\nc{\beqa}{\begin{eqnarray}}
\nc{\eeqa}{\end{eqnarray}}
\nc{\lbar}[1]{\overline{#1}}
\nc{\ketbra}[2]{|#1\rangle\!\langle#2|}
\nc{\braket}[2]{\langle#1|#2\rangle}
\nc{\proj}[1]{| #1\rangle\!\langle #1 |}
\nc{\avg}[1]{\langle#1\rangle}
\nc{\Rank}{\operatorname{Rank}}
\nc{\smfrac}[2]{\mbox{$\frac{#1}{#2}$}}
\nc{\Tr}{\operatorname{Tr}}
\nc{\id}{\operatorname{id}}
\nc{\1}{\openone}
\nc{\ox}{\otimes}
\nc{\dg}{\dagger}
\nc{\dn}{\downarrow}
\nc{\cA}{{\cal A}}
\nc{\cB}{{\cal B}}
\nc{\cC}{{\cal C}}
\nc{\cD}{{\cal D}}
\nc{\cE}{{\cal E}}
\nc{\cF}{{\cal F}}
\nc{\cG}{{\cal G}}
\nc{\cH}{{\cal H}}
\nc{\cI}{{\cal I}}
\nc{\cJ}{{\cal J}}
\nc{\cK}{{\cal K}}
\nc{\cL}{{\cal L}}
\nc{\cM}{{\cal M}}
\nc{\cN}{{\cal N}}
\nc{\cO}{{\cal O}}
\nc{\cP}{{\cal P}}
\nc{\cR}{{\cal R}}
\nc{\cS}{{\cal S}}
\nc{\cT}{{\cal T}}
\nc{\cX}{{\cal X}}
\nc{\cY}{{\cal Y}}
\nc{\cZ}{{\cal Z}}
\nc{\var}{\operatorname{var}}
\nc{\rar}{\rightarrow}
\nc{\lrar}{\longrightarrow}
\nc{\polylog}{\operatorname{polylog}}

\def\e{\epsilon}

\nc{\RR}{{{\mathbb R}}}
\nc{\CC}{{{\mathbb C}}}
\nc{\FF}{{{\mathbb F}}}
\nc{\NN}{{{\mathbb N}}}
\nc{\ZZ}{{{\mathbb Z}}}
\nc{\PP}{{{\mathbb P}}}
\nc{\QQ}{{{\mathbb Q}}}
\nc{\UU}{{{\mathbb U}}}
\nc{\EE}{{{\mathbb E}}}
\nc{\Icoh}{{I^{\rm coh}}}
\nc{\Qca}{{Q_{\rm ss}}}
\nc{\Qcaa}{{Q^{(1)}_{\rm ss}}}
\nc{\Dcaa}{{D^{(1)}_{{\rm ss}\rightarrow}}}
\nc{\Dca}{{D_{{\rm ss}\rightarrow}}}

\nc{\be}{\begin{equation}}
\nc{\ee}{{\end{equation}}}
\nc{\bea}{\begin{eqnarray}}
\nc{\eea}{\end{eqnarray}}
\nc{\<}{\langle}
\rnc{\>}{\rangle}
\nc{\Hom}[2]{\mbox{Hom}(\CC^{#1},\CC^{#2})}
\nc{\rU}{\mbox{U}}

\begin{document}

\title{High performance single-error-correcting quantum codes for amplitude damping}

\author{Peter W. Shor, Graeme Smith, John A. Smolin, Bei Zeng
\thanks{PW Shor is with the Department of Mathematics, Massachusetts Institute of Technology, Cambridge, MA 02139, USA}
\thanks{G. Smith and JA Smolin are with the IBM T.J. Watson Research Center, Yorktown Heights, NY 10598, USA}
\thanks{B. Zeng is with the Department of Physics, Massachusetts Institute of Technology, Cambridge, MA 02139, USA and was with the IBM T.J. Watson Research Center, Yorktown Heights, NY 10598, USA}}
\date{\today}

\maketitle
\begin{abstract}
We construct families of high performance quantum amplitude damping
codes.  All of our codes are nonadditive and most modestly outperform
the best possible additive codes in terms of encoded dimension.  One
family is built from nonlinear error-correcting codes for
classical asymmetric channels, with which we systematically construct
quantum amplitude damping codes with parameters better than any prior
construction known for any block length $n\geq 8$ except $n=2^r-1$. We
generalize this construction to employ classical codes over $GF(3)$
with which we numerically obtain better performing codes up to
length $14$.  Because the resulting codes are of the codeword
stabilized (CWS) type, easy encoding and decoding circuits are
available.

\end{abstract}



\section{Introduction}
Quantum computers offer the potential to solve certain classes of
problems that appear to be intractable on a classical machine.  For example,
they allow for efficient prime factorization \cite{shor}, breaking
modern public-key cryptography systems based on the assumption that
factorization is hard.  Quantum computers may also be useful for
simulating quantum systems \cite{Feynman,lloyd}.

However, quantum computers are particularly subject to the deleterious
effects of noise and decoherence.  It was thought, for a time, that
quantum error-correction would be precluded by the no cloning theorem
\cite{nocloning} which seems to rule out redundancy as usually
employed in error correction.  The discovery of quantum
error-correcting codes \cite{shorECC,Steane0} that allow for fault-tolerant
quantum computing \cite{shorfault} significantly bolstered the hopes of building
practical quantum computers.

For the most part, people have concentrated on dealing with the worst
case---arbitrary (though hopefully small) noise.  This turns out to be
equivalent to correcting Pauli-type errors, $\X={ 0\ 1 \choose
  1\ 0},\Y={0\ -i \choose i\ \ 0},\Z={1 \ \ 0\choose 0\ -\!1}$, acting
on a bounded-weight subset of the qubits in the code.
Since the Pauli operators form a basis of $2\times 2$ matrices, a code that can
correct all Pauli errors can in also protect against any general qubit
noise \cite{bigpaper,chiara}.
 
However, as first demonstrated by Leung et al. \cite{Debbie},
designing a code for a particular type of noise can result in codes
with better performance. In practice the types of noise seen are
likely to be unbalanced between amplitude ($\X$-type) errors and phase
($\Z$-type) errors, and recently a lot of attention has been put into
designing codes for this situation and in studying their fault
tolerance properties \cite{Lev} \cite{Panos} \cite{Evans}
\cite{Peter1}.

In this paper, we will focus on {\em amplitude damping} noise, another
type of noise seen in realistic settings.  Amplitude damping noise is
asymmetric, with some chance of turning a spin up $\ket{1}$ qubit into
a spin down $\ket{0}$ state but never transforming $\ket{0}$ to
$\ket{1}$.  This models, for example, photon loss in an optical fiber:
A photon in the fiber may leak out or absorbed by atoms in the fiber,
but to good approximation photons do not spontaneously appear in the
fiber.  Several people have considered this type of noise
\cite{Debbie,Peter1,ChuangLeungYamamoto} but there is no systematic
method for constructing such codes.  In general it is a difficult
problem to design codes for any particular noise model.

In this paper we present a method for finding families of codes
correcting one amplitude-damping error.  We begin with an ansatz
relating a restricted type of amplitude-damping code to classical
codes for the binary asymmetric (or $Z$-) channel.  The $Z$-channel is
the classical channel that takes 1 to 0 with some probability, but never
vice versa\footnote{Not to be confused with quantum $\Z$ errors, the
channel takes its name from its diagram resembling the letter 'Z.'  See
Figure \protect\ref{fig:channel}.}.
 The amplitude damping channel is its natural quantum
generalization.  The problem of designing codes for
the amplitude damping channel is thus reduced to a finding classical 
codes for the $Z$-channel, subject to a constraint.  This lets us 
carry over many known results from classical coding theory.

We further simplify the problem by using a novel mapping between binary and 
ternary codes.  This allows us to find quantum amplitude-damping codes by 
studying ternary codes on a greatly reduced search space.

The rest of the paper is organized as follows.  In section
\ref{prelims} we describe quantum channels and the quantum
error-correction conditions.  In section \ref{AD} we define what it
means to correct amplitude damping errors and show how they relate to
classical symmetric codes.  In section \ref{symm} we show how a
particular class of amplitude-damping codes arises from classical
codes for the asymmetric channel, and give some new codes based on
powerful extant results on classical $Z$-channel codes \cite{CR}.  In
section \ref{GF3} we define a mapping from binary to ternary codes
(and back) and use this to construct new and better amplitude damping
codes.  Finally, in section \ref{summary} we summarize our results and
give a table of the best amplitude-damping codes and how they compare
to previous work.

\section{Preliminaries\label{prelims}}
Pure quantum states are represented by vectors in a complex
vector space.  We will be concerned with finite-dimensional
systems.  The simplest quantum system (called a qubit) can be
described by an element of $\CC^2$, and $n$
qubits together are described by an elements of 
$\CC^2 \otimes \ldots \otimes \CC^2=  (\CC^2)^{\otimes n}.$
Such pure states are always chosen to be normalized to unity.
More generally a quantum system can be described by a density matrix, 
a trace one linear operator from $(\CC^2)^{\otimes n}$ to $(\CC^2)^{\otimes n}$,
usually denoted $\rho$.

The most general physical transformations allowed by the quantum mechanics are
completely positive, trace preserving linear maps which can be represented
by the Kraus decomposition:
\begin{equation}
\cN(\rho)=\sum_k A_k \rho A_k^\dag\ {\rm where\ }\sum_k A_k^\dag A_k = \iden .
\end{equation}
For example the the Kraus operators for the depolarizing channel, the
natural quantum analogue of the binary symmetric channel, are the Pauli
matrices.
  The Kraus operators for the amplitude damping channel with damping rate $\epsilon$ are
\begin{equation}
A_0=\begin{pmatrix} 1 & 0 \\0 &
\sqrt{1-\epsilon} \end{pmatrix}{\rm\ and\ } A_1=\begin{pmatrix} 0 &
\sqrt{\epsilon} \\0 & 0 \end{pmatrix} .
\end{equation}

A quantum error correcting code is subspace of $(\CC^2)^{\otimes n}$ which
is resilient to some set of errors acting on the individual qubits such that
all states in that subspace can be recovered.  For a $d$-dimensional
codespace spanned 
by the orthonormal set $\ket{\psi_i}$, $i=1 \ldots d$ and a set of errors $\cE$
there is a physical operation correcting all elements $E_\mu \in \cE$ if the
error correction conditions \cite{KnillLF,bigpaper} are satisfied:
\begin{equation}
\forall_{ij,\mu\nu}\ \ \bra{\psi_i} E_\mu^\dag E_\nu \ket{\psi_j} = C_{\mu\nu}\delta_{ij}, 
\end{equation}
where $C_{\mu \nu}$ depends only on $\mu$ and $\nu$.

\section{Correcting amplitude damping \label{AD}} 

For small $\epsilon$, we would like to correct the leading order errors that occur during amplitude damping.
Letting $A=\X+i\Y$, $B=I-\Z$, we have
\begin{equation}
A_1=\frac{\sqrt{\epsilon}}{2}A,\ \  A_0=I-\frac{\epsilon}{4}(I-\Z)+O(\e^2).
\end{equation}
It can be shown if we wish to improve fidelity through an amplitude
damping channel from $1-\epsilon$ to $1-\epsilon^t$ it is sufficient
to satisfy the error-detection conditions for $2t$ $A$ errors and $t$
$\Z$ errors.  We will say the such a code corrects $t$ amplitude damping
errors since it improves the fidelity, to leading order, just as much
as a true $t$-error-correcting code would for the same channel.  We
will use the notation $\lfloor\lfloor n,K,t \rfloor\rfloor$ to mean an
$n$-qubit code protecting a $K$-dimensional space and correcting $t$
amplitude damping errors, sometimes referring to this as a $t$-AD code.
Our notation descends from the traditional coding-theory notation of $[n,k,d]$
to mean an $n$-bit classical code of distance $d$ protecting $k$ bits and
$[[n,k,d]]$ to mean an $n$-qubit quantum code of distance $d$ protecting 
$k$ qubits.  Note that our AD notation uses $K$ as the full dimensions of the 
protected space, {\em not} $k$, the $\log$ of the dimension.  This is
in preparation for the codes we will design which do not protected an
integral number of qubits.

Since the amplitude damping channel is not a Pauli channel
the usual tools for designing quantum codes cannot be directly used.
One possible approach would be to design CSS
\cite{shorECC,Steane0,Steane2} codes with different $\X,\Z$ distances
\cite{Steane}. For the particular case of single-error-correcting AD
code, we then would like to have CSS code of $\X$ distance $3$
(correcting a single $\X$ error) and $\Z$ distance $2$ (detecting a
single $\Z$ error). Gottesman gives a construction of this kind of CSS
code in Chapter 8.7 of \cite{Daniel}. We summarize his result as
follows:
\begin{theorem}
\label{th:ADCSS}
If there exists a binary $[n,k,3]$ classical code ${\mathcal C}$ and ${\mathbf 1}$ (the all $1$ string of length $n$) is in the dual code of ${\mathcal C}$, then there exists an $\lfloor\lfloor n,2^{k-1},1\rfloor\rfloor$ code.
\end{theorem}

These codes indeed have better performance than codes designed for
depolarizing channels. For instance, a $\lfloor\lfloor 7,2^3,1 \rfloor\rfloor$ 
exists while only $[[7,1,3]]$ single-error-correcting stabilizer codes exist for the depolarizing channel. In general, the classical Hamming bound for $[n,k,3]$ codes gives $k\leq n-\log(n+1)$, which gives a bound for $[[n,k]]$ single-error-correcting AD codes constructed by Theorem \ref{th:ADCSS}, {\em i.e.}
\begin{equation}
k\leq n-1-\log(n+1),
\end{equation}
while the quantum Hamming bound ({\em cf.} \cite{Daniel}) gives
\begin{equation}
k\leq n-\log(3n+1)
\end{equation}
for $[[n,k,3]]$ stabilizer codes for the depolarizing channel.

However, one expects that these codes cannot be
optimal; since we only need to correct $\X+i\Y$,
correcting both $\X$ and $\Y$ is excessive and would seem to lead to
inefficient codes. Fletcher et al. took the first step toward making AD codes based on the non-Pauli error model, {\em i.e.} codes correcting $\X+i\Y$ error, not both $\X$ and $\Y$ errors \cite{Peter1}. Their codes are stabilizer codes with parameters $[[2n,n-1]]$ and correct a single amplitude damping error. Later another work \cite{Peter2} took a further step toward making AD codes correcting $\X+i\Y$ error. These works constructed some nonadditive codes correcting a single amplitude damping error, and via numerical search for short block length found AD codes with better performance than codes given by the CSS construction of Theorem \ref{th:ADCSS}.

The construction of \cite{Peter2} consists of codewords $\ket{\psi_u}$ of the self-complementary format \cite{family}, which is
\begin{equation}
\ket{\psi_u}=\frac{1}{\sqrt{2}}\left(|u\rangle+|\bar{u}\rangle\right),
\label{SC}
\end{equation}
where $u$ is a binary string of length $n$ and $\bar{u}={\mathbf 1}\oplus u$. 

As observed in \cite{family}, which focused on nonadditive single-error-detecting codes, codes consisting of codewords given by Eq. (\ref{SC}) automatically detect a single $\Z$ error, so we have, as shown in \cite{Peter2}:
\begin{theorem}
\label{th:ADSC}
A self-complementary code corrects a single amplitude damping error if and only if no confusion arises assuming the decay occurs at no more than one qubit.
\end{theorem}

We will take the above observation as a starting point for making
amplitude damping codes, by choosing classical self-complimentary codes
which correct single errors arising from the classical asymmetric channel 
(or $Z$-channel).

\section{Systematic construction from classical asymmetric codes \label{symm}}

\begin{figure}[h!]
\centering
\includegraphics[scale=0.4,angle=0]{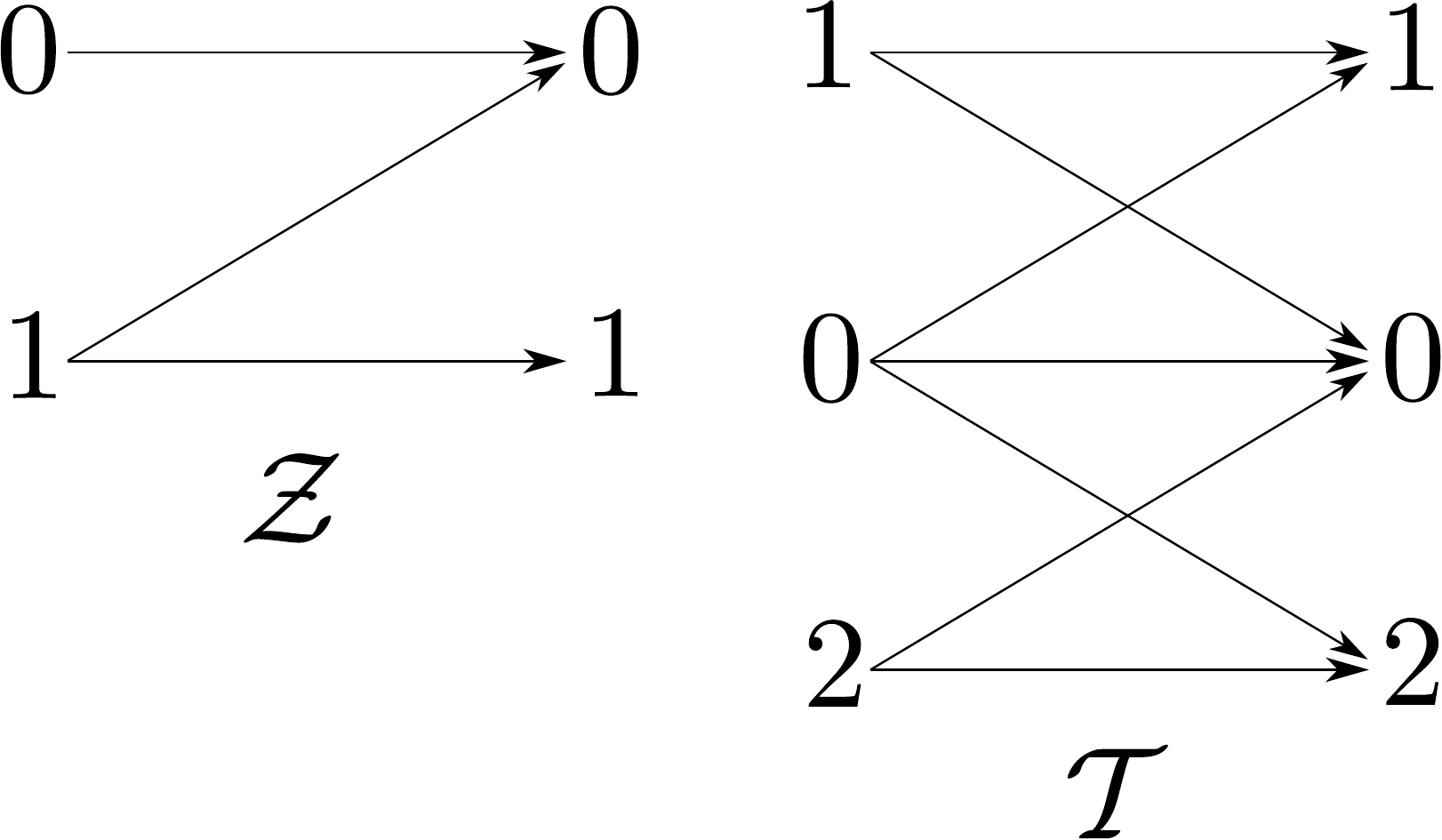}
\caption{The binary asymmetric channel ${\mathcal Z}$ and the ternary
channel ${\mathcal T}$.}
\label{fig:channel}
\end{figure}

Now we would like to relate the self-complementary construction to
classical error correcting codes for the asymmetric channel. Before doing
that we first briefly review the classical theory of those codes.
\begin{definition}
The \textbf{binary asymmetric channel} (denoted by ${\mathcal Z}$ in
Fig. \ref{fig:channel}) is the channel with $\{0,1\}$ as input and output
alphabets, where the crossover $1\rightarrow 0$ occurs with positive
probability $p$, whereas the crossover $1\rightarrow 0$ never occurs.
\end{definition}
We will call a classical code that protects against one error in the
binary asymmetric channel ${\mathcal Z}$ a {\em $1$-code} and use the
notation $\lfloor n,K,t\rfloor$ analogous to our notation for the
quantum amplitude damping code.

We can then formalize our observation as:
\begin{theorem}
\label{th:ADtoAS}
If ${\mathcal C}$ is a classical $\lfloor n,K,1 \rfloor$ code and $\forall u\in
{\mathcal C}$, $\bar{u}\in {\mathcal C}$, then $Q=\{|u\rangle+|\bar{u}\rangle,\ u\in {\mathcal C}\}$ is
a single-error correcting amplitude damping code, $\lfloor\lfloor
n,K/2,1\rfloor\rfloor$.
\end{theorem}

This theorem is almost a direct corollary of Theorem \ref{th:ADSC} so we omit a detailed proof. The main idea is that a
 classical code ${\mathcal C}$ that contains both $u$ and $\bar{u}$ takes care of correcting amplitude damping errors 
while the self-complementary form of $\ket{\psi_u}$ takes care of detecting the phase errors. 
And the size of the quantum code $Q$ is of course $K=|{\mathcal C}|/2$. This theorem allows us 
to use any classical self-complimentary $1$-code to construct self-complementary amplitude damping codes. The question that 
remains is how to find classical self-complimentary $1$-codes.

Varshamov showed almost all linear codes that are able to
correct $t$ asymmetric errors are also able to correct $t$ symmetric
errors \cite{Varshamov1}. Therefore, to go beyond $t$-symmetric-error
correcting codes, we will look to non-linear constructions. Note that
the quantum codes we construct from these non-linear codes are 
codeword stabilized codes, so these nonlinear classical codes will typically
result in nonadditive quantum codes \cite{CWS}.

\subsection{Constantin-Rao Codes}
Constantin-Rao ({\em CR}) Codes
\cite{CR} are the best known nonlinear $1$-codes. These beat the best symmetric
single-error-correcting codes for all $n\neq 2^r-1$. 
An $n$-bit CR codes is constructed based on an abelian group $G$ of
size $n+1$. The group operation is written as `$+$' for abelian
groups.
\begin{definition}
The Constantin-Rao code ${\mathcal C}_g\ \forall g\in G$  is given by
\begin{equation}
{\mathcal C}_g=(\{(x_1,x_2,...,x_n)|\sum_{i=1}^n x_ig_i=g\ \mod\ n+1\}),
\end{equation}
where $x_i\in\{0,1\}$ and $g_1,g_2,...,g_n$ are the non-identity elements of $G$.
\end{definition}

The cardinality of ${\mathcal C}_g$ is lower bounded by
\begin{equation}
|{\mathcal C}_g|\geq \frac{2^n}{n+1}
\end{equation}
for some $g\in G$.

Let $o(g)$ be the order of $g$, then it is known
\begin{equation}
|{\mathcal C}_0|\geq |{\mathcal C}_g|,
\end{equation}
with equality if and only if $o(g)$ is a power of $2$.

For a given nonprime $n+1$, there may be many abelian groups of size $n+1$. If the group $G$ is a cyclic group of order $n+1$, then the corresponding codes are called Varshamov-Tenengol'ts codes \cite{Varshamov2}.  It is known that the largest Constantin-Rao code of length $n$ is the code ${\mathcal C}_0$ based on the group $G=\bigoplus_{p|n+1}\bigoplus_{i=1}^{n_p}\mathbb{Z}_p$, where $n+1=\Pi_{p|n+1}p^{n_p}$ \cite{Klove}.

An exact expression for the size of a CR code
based on the group properties is known, and a basic result is that for
any group $G$ and any group element $g$, $|{\mathcal C}_g|$ has size
approximately $\frac{2^n}{n+1}$ (for a review, see \cite{Klove}).
Note $\frac{2^n}{n+1}$ is the Hamming bound for $1$-error correcting
codes over the binary symmetric channel. Thus, CR codes provide excellent performance compared to symmetric codes
and, indeed, outperform the best known symmetric codes for all block-lengths but $n = 2^r-1$.

\subsection{Amplitude damping codes from Constantin-Rao codes}

To build quantum codes from ${\mathcal C}_g$, we need to find CR codes which are
self-complimentary (and preferably large).  We will show these exist
for all $n > 1$.
\begin{fact}
\label{fact:even}
For even $n$, the Constantin-Rao code ${\mathcal C}_0$ is self-complementary.
\end{fact}
This is based on a simple observation that all the nonzero group elements add up to zero for any abelian group of even size. 

The case of odd lengths $n$ is more complicated. We first consider the case where $n=4k+3$. Recall that the largest Constantin-Rao code of length $n$ is the code ${\mathcal C}_0$ based on the group $G=\bigoplus_{p|n+1}\bigoplus_{i=1}^{n_p}\mathbb{Z}_p$, where $N=\Pi_{p|n+1}p^{n_p}$. Then further note that for an abelian group $\mathbb{Z}_2\oplus\mathbb{Z}_2\oplus\mathbb{G}$, where the group $\mathbb{G}$ is of odd size,  all the nonzero group elements add up to zero. This leads to the following
\begin{fact}
For $n=4k+3$, the Constantin-Rao code ${\mathcal C}_0$ of the maximal cardinality is self-complementary.
\end{fact}

Since $|{\mathcal C}_0|\geq |{\mathcal C}_g|\geq \frac{2^n}{n+1}$, AD codes constructed from Fact \ref{fact:even} and Fact \ref{fact:4k3} outperform the CSS AD codes of even length and odd length $n=4k+3$ constructed by Theorem \ref{th:ADCSS}.  

Note we also have
\begin{fact}
\label{fact:4k3}
For $n=4k+3$, the Varshamov-Tenengol'ts code ${\mathcal V}_\frac{n+1}{4}$ of the maximal cardinality is self-complementary.
\end{fact}
 
The case for $n=4k+1$ is more tricky. We cannot directly get a self-complementary code of length $n$ from some Constantin-Rao codes ${\mathcal C}_g$ of the same length $n$. But instead we can construct self-complementary AD codes of length $n$ from the Varshamov-Tenengol'ts codes ${\mathcal V}_g$ of length $n+1$.
\begin{fact}
\label{fact:4k1}
For $n=4k+1$, the shortened Varshamov-Tenengol'ts code ${\mathcal V}'_{\frac{n+2-r}{2}}$ obtained by deleting an odd coordinate $r$ from Varshamov-Tenengol'ts code ${\mathcal V}_{\frac{n+2-r}{2}}$ of length $n+1$ is self-complementary.
\end{fact}
The codewords of this shortened Constantin-Rao code are given by
\begin{equation}
\sum_{i=1,i\neq r}^{n+1} ix_i=\frac{n+2-r}{2} \mod\ n+2.
\label{nequals4kplus1}
\end{equation}
Since $\sum_{i=1,i\neq r}^{n+1} i \mod\ n+2 = n+2-r$, for any set of $x_i$s we have
\begin{equation}
\sum_{i=1,i\neq r}^{n+1} ix_i +  i\bar{x}_i \mod\ n+2 = n+2-r\ 
\end{equation}
where $x_i \in \{0,1\}$ and $\bar{x}_i=1 \oplus x_i$.
If the $x_i$s satisfy (\ref{nequals4kplus1}) then so do the $\bar{x}_i$s.  
Therefore ${\mathcal V}'_{\frac{n+2-r}{2}}$ is self-complementary. 

It is known that the size of these shortened Varshamov-Tenengol'ts codes are approximately $\frac{2^n}{n+2}$ \cite{Klove}. But we know that the size of binary symmetric codes for length $n=4k+1$ is upper bounded by $\frac{2^n}{n+2}$ \cite{MacWilliams}, so the construction of AD codes given by Fact \ref{fact:4k1} also outperforms the CSS AD codes of length $n=4k+1$ constructed by Theorem \ref{th:ADCSS}.

\begin{example}
\label{eg:AD816}
For $n=8$, choose the abelian group of size $n+1=9$ be $\mathbb{Z}_3\oplus\mathbb{Z}_3$. The codewords of the Constantin-Rao code ${\mathcal C}_0$ are given by a linear code ${\mathcal C}_1$ generated by 
\begin{equation}
\{00000011, 00001100, 00110000\}; 
\end{equation}
and four pairs $P_i$ (i=1\ldots 4): 
\begin{eqnarray}
{\mathcal P}_1&=&\{10100001, 10101101\},\nonumber\\
{\mathcal P}_2&=&\{10000110, 10110110\},\nonumber\\ 
{\mathcal P}_3&=&\{01100100, 01100111\},\nonumber\\ 
{\mathcal P}_4&=&\{00101010, 11101010\};
\end{eqnarray} 
and all the complements of $\bigcup_{i=1}^{4}{\mathcal P}_i\bigcup {\mathcal C}$.

The weight distribution of this code is given by (for definition of weight distribution, see \cite{Shor-laflamme-1997, Rains-weight-1998})
$A_0=1; A_1=0; A_2=1/4; A_3=0; A_4=9/2; A_5=0; A_6=9/4; A_7=0; A_8=8.$ 
Some of them are non-integers, so this code is nonadditive.

The size of the quantum code is $16$, so this is a 
$\lfloor\lfloor 8,2^4,1\rfloor\rfloor$ code. Note the CSS AD code constructed by Theorem \ref{th:ADCSS} for $n=8$ gives parameters $\lfloor\lfloor 8,2^3,1\rfloor\rfloor$. And the best single-error-correcting stabilizer code for the depolarizing channel is $[[8,3,3]]$. Therefore, this nonadditive AD code encodes one more logical qubit than the best known stabilizer code with the same length and is capable of correcting a single amplitude damping error.
\end{example}

For short block length ($\leq 16$), a comparison of the code dimensions given by this Constantin-Rao construction with other constructions will be listed in Table \ref{table:ADcodes} in Sec. \ref{sec:summary}. One can see that this Constantin-Rao construction outperforms all the other constructions apart from the $GF(3)$ construction given in Sec.~\ref{GF3}. However, since the $GF(3)$ construction is not systematic (those codes given by the $GF(3)$ construction in Table \ref{table:ADcodes} are found by numerical search), this Constantin-Rao construction is the best known systematic construction for single-error-correcting AD codes.

\section{The $GF(3)$ construction and the ternarization map \label{GF3}}

We will begin by defining a channel ${\mathcal T}$ which acts on a three
letter alphabet and find ternary codes on this channel.  We will
then show that such codes are related to binary codes for the asymmetric
channel and since the binary codes will be self-complimentary by construction
that they will yield quantum amplitude damping codes as well.

\subsection{The ternarization map}

\begin{definition}
The \textbf{ternary channel} (denoted ${\mathcal T}$ in the figure) has
$\{0,1,2\}$ as input and output alphabets, where the crossovers
$0\rightarrow 0$, 
$0\rightarrow 1$, 
$0\rightarrow 2$, 
$1\rightarrow 0$, 
$1\rightarrow 1$, 
$2\rightarrow 0$, and
$2\rightarrow 2$ all occur with nonzero probability, but 
$1\rightarrow 2$ and  
$2\rightarrow 1$ never occur.
\end{definition}

We define a map that takes pairs of binary coordinates into a single ternary
coordinate.  There are four possible values of binary pairs, and only
three ternary coordinates, so it cannot be one-to-one.

\begin{definition}
The ternarization map $\tilde{\mathfrak{S}}: \mathbb{F}^2_2\rightarrow \mathbb{F}_3$ is defined by:
\begin{equation}
\tilde{\mathfrak{S}}:\ \{00,11\}\rightarrow 0,\ 01\rightarrow 1,\ 10\rightarrow 2.
\end{equation}
\end{definition}

This is not a one to one map. So the inverse map needs to be specified carefully, that is, a ternary symbol $0$ after the inverse map gives two binary codewords $00$ and $11$.
\begin{definition}
The map $\mathfrak{S}: \mathbb{F}_3\rightarrow \mathbb{F}^2_2$ is defined by:
\begin{equation}
\mathfrak{S}:\ 0\rightarrow \{00,11\},\ 1\rightarrow 01,\ 2\rightarrow 10.
\end{equation}
\end{definition}

For a binary code of length $n=2m$, by choosing a pairing of coordinates, the map $\tilde{\mathfrak{S}}^m: \mathbb{F}^{2m}_{2}\rightarrow \mathbb{F}_3^m$ then takes a given binary code of length $2m$ to a ternary code of length $m$.
\begin{example}
The optimal $1$-code ${\mathcal C}^{(4)}$ of length $n=4$ and dimension $4$ has four codewords $\{0000,1100,0011,1111\}$. By pairing coordinates $\{1,2\}$ and $\{3,4\}$, the ternary image under $\tilde{\mathfrak{S}}^2$ is then $\{00\}$.
\end{example}

On the other hand,   $\mathfrak{S}^m: \mathbb{F}_3^m\rightarrow \mathbb{F}^{2m}_2$ takes a given ternary code of length $m$ to a binary code of length $2m$. 
\begin{example}
\label{eg:ternary423}
By starting from the linear ternary code $[4,2,3]_3$, with generators $\{0111,1012\}$, we get the binary image code ${\mathcal C}^{(8)}$ under $\mathfrak{S}^4$: 
\begin{equation}
\begin{array}{llll}
00000000 & 00000011 & 00001100 & 00001111 \\
00110000 & 00110011 & 00111100 & 00111111 \\
11000000 & 11000011 & 11001100 & 11001111 \\
11110000 & 11110011 & 11111100 & 11111111 \\
00010101 & 00101010 & 11010101 & 11101010\\
01000110 & 10001001 & 01110110 & 10111001 \\
01011000 & 10100100 & 01011011 & 10100111 \\
10010001 & 01100010 & 10011101 & 01101110 \\
\end{array}
\end{equation}
which is of dimension $32$ and corrects one asymmetric error. Note this gives exactly the same binary $1$-code as the one given in Example \ref{eg:AD816}, which is the Constantin-Rao code ${\mathcal C}_0$ of length $n=8$ constructed from the group  $\mathbb{Z}_3\oplus\mathbb{Z}_3$. This example hints at some relationship between the $GF(3)$ construction and the  Constantin-Rao codes. 
\label{Eg832}
\end{example}

\subsection{The $GF(3)$ construction for  $1$-codes
\label{sec:GF3classic}}

\subsubsection{Even block length}
Example \ref{Eg832} suggests that good $1$-codes may be obtained from ternary codes under the map $\mathfrak{S}^m$. We would like to know the general conditions under which a ternary code gives a $1$-code via the map $\mathfrak{S}^m$. The main result of this section states that any single-error-correcting code for the ternary channel ${\mathcal T}$ gives a $1$-code under the map $\mathfrak{S}^m$ \cite{gf3classical}.

It will be useful in what follows to define an asymmetric distance between two codewords: 
\begin{definition}
Letting $N(\mathbf{x},\mathbf{y})=\#\{i|x_i=0\ \text{and}\ y_i=1\}$, we define the
{\em asymmetric distance} between $\mathbf{x}$ and $\mathbf{y}$ as
\begin{equation}
\Delta(\mathbf{x},\mathbf{y}):=\max\{N(\mathbf{x},\mathbf{y}),N(\mathbf{y},\mathbf{x})\}.
\end{equation}
\end{definition}
It is easy to see that a set of codewords with minimum asymmetric distance $2$ is a $1$-code.

\begin{theorem}
If ${\mathcal C}'$ is a single-error-correcting ternary code for the channel ${\mathcal T}$ of length $m$, then ${\mathcal C}=\mathfrak{S}^{m}({\mathcal C}')$ is a $1$-code of length $2m$.
\label{ChannelT}
\end{theorem}
\textbf{Proof} 
For any two ternary codewords  $\mathbf{c}'_1,\mathbf{c}'_2\in \mathcal{C}'$, we need to show that the asymmetric distance between 
$\mathfrak{S}^m(\mathbf{c}'_1)$ and $\mathfrak{S}^m(\mathbf{c}'_2)$ is at least two. 

First, we cover the case when $\mathbf{c}'_1=\mathbf{c}'_2$.  Distinct binary codewords may arise from the same ternary codeword due to the two different actions of $\mathfrak{S}$ on $0$.  Such codewords have $\Delta \ge 2$ since $\Delta(00,11)=2$.

Next, if the Hamming distance between $\mathbf{c}'_1$ and $\mathbf{c}'_2$ is three, then the distance between $\mathfrak{S}^m(\mathbf{c}'_1)$ and $\mathfrak{S}^m(\mathbf{c}'_2)$ is also three since $\Delta(00,01), \Delta(11,01),\Delta(00,10),\Delta(00,01),$ and $\Delta(01,10)$ are all one and three such $\Delta$s occur.

Finally, the following Hamming distance two pairs are allowed in a single-error-correcting ternary code for $\mathcal T$:
\begin{equation}
\begin{array}{lllll}
01,22 & 10,22 & 01,12 & 10,21 & 02,11\\
20,11 & 02,21 & 20,12 & 11,22 & 12,21 \\
\end{array}
\label{pairs}
\end{equation}
It is straightforward to verify that $\mathfrak{S}$ on these pairs also results in binary codes with $\Delta \ge 2$.
$\square$

The following corollary is straightforward.
\begin{corollary}
If ${\mathcal C}'$ is a linear $[n,k,3]_3$ code (the subscript indicates that 
the code is over a three-letter alphabet rather than a binary alphabet), then $\mathfrak{S}^{m}({\mathcal C}')$ is a $1$-code of length $2m$.
\label{linear}
\end{corollary}

\subsubsection{Odd block length
\label{oddonecode}}
Theorem \ref{ChannelT} only works for designing $1$-codes of even length. Now we generalize this construction to the odd length situation, starting from `adding a bit' to the ternary code \cite{gf3classical}.
\begin{definition} 
We call a code acting on $\mathbb{F}_2 \times \mathbb{F}_3^m$ a {\em generalized ternary code} of length $m+1$.  We further adopt the conventions that
$\mathfrak{S}^m({\mathcal C}')$ gives a $(2m+1)$-bit binary code by acting on the $m$ trits of a generalized ternary code ${\mathcal C}'$ and $\tilde{\mathfrak{S}}^{2m}({\mathcal C})$ when ${\mathcal C}$ has length $2m+1$ gives a generalized ternary code by acting on the last $2m$ bits of ${\mathcal C}$.
\end{definition}
\begin{theorem}

If ${\mathcal C}'$ is a single-error-correcting generalized ternary code for the channel ${\mathcal Z}\times{\mathcal T}^{m}$ of length $m+1$, then ${\mathcal C}=\mathfrak{S}^{m}({\mathcal C}')$ is a $1$-code of length $2m+1$.
\label{ChannelZT}
\end{theorem}

\noindent \textbf{Proof} 

As in the proof of Theorem \ref{ChannelT} we need to 
show that for any two codewords
$\mathbf{c}'_1,\mathbf{c}'_2\in \mathcal{C}'$, we need to show that
the asymmetric distance between $\mathfrak{S}^m(\mathbf{c}'_1)$ and
$\mathfrak{S}^m(\mathbf{c}'_2)$ is at least two.  If the Hamming distance
between codewords on {\em just the ternary} part of the code is at least two, then the situation reduces to the previous proof.

We need only worry about the case where the Hamming distance between
$\mathbf{c}'_1$ and $\mathbf{c}'_2$ is two, and one of the differences
in on the binary coordinate.  Assume the first coordinate is a bit and
the second is a trit, then since ${\mathcal C}'$ is a
single-error-correcting generalized ternary code the only allowed
pairs are $01,12$; and $ 12,11$. The corresponding images of each pair
under $\mathfrak{S}^m$ give binary codewords of asymmetric distance
$\Delta=2$.  $\square$

To illustrate this generalized ternary construction, let us look at the following example.
\begin{example}
The code $\{0000,0111,0222,1012,1120,1201\}$ corrects a single error from the channel ${\mathcal Z}\times{\mathcal T}^{3}$. Under the map $\mathfrak{S}^3$ it gives the binary code
\begin{equation}
\begin{array}{llll}
0000000 & 0000011 & 0001100 & 0001111 \\
0110000 & 0110011 & 0111100 & 0111111 \\
0010101 & 0101010 & 1000110 & 1110110 \\
1011000 & 1011011 & 1100001 & 1101101 \\
\end{array}
\end{equation}
which is a binary code of length $7$, dimension $16$ which corrects one asymmetric error.
\label{gt7}
\end{example}

The following corollary is straightforward, but gives the most general situation of the ternary construction.
\begin{corollary}
If ${\mathcal C}'$ is a ternary single error correcting code of channel ${\mathcal Z}^{{m_1}}\times{\mathcal T}^{{m_2}}$ of length $m_1+m_2$, then ${\mathcal C}=\mathfrak{S}^{m_2}({\mathcal C}')$ is a $1$-code of length $m_1+2m_2$.
\label{general}
\end{corollary}

\subsection{The $GF(3)$ construction for AD codes}

\subsubsection{Even block length}

We first examine under which conditions the image of a ternary code under $\mathfrak{S}$ could be self-complementary.
\begin{definition}
A ternary code ${\mathcal C}'$ is self-complementary if for any $\mathbf{c}'\in {\mathcal C}'$, $\bar{\mathbf{c}}'\in {\mathcal C}'$, where $\bar{\mathbf{c}}=({\mathbf 3}\ominus \mathbf{c})\ \mod 3$ (${\mathbf 3}=33\ldots 3$, {\em i.e.} the all `3' string).
\end{definition}
\begin{example}
The ternary code ${\mathcal C}'=\{000,111,222\}$ is self-complementary. For $111\in {\mathcal C}'$, $\overline{111}=333\ominus 111=222$.
\end{example}

\begin{definition}
We say that binary code ${\mathcal C}$ of even length $n=2m$ has {\em ternary form} if $\mathfrak{S}^{m}(\tilde{\mathfrak{S}}^m({\mathcal C}))={\mathcal C}$.
\end{definition}

The properties of $\mathfrak{S}$ gives the following
\begin{fact}
\label{fact:tecomp}
If a ternary code ${\mathcal C}'$ of length $m$ is self-complementary, then its binary image under $\mathfrak{S}$,  ${\mathcal C}=\mathfrak{S}^m({\mathcal C}')$,  is self-complementary. On the other hand, if a binary code ${\mathcal C}$ of length $2m$ is of ternary form and is self-complementary, then its ternary image $\tilde{\mathfrak{S}}^{2m}({\mathcal C})$ is self-complementary.
\end{fact}

To use Fact \ref{fact:tecomp} to construct good single-error-correcting AD codes for even block length, first recall Example \ref{eg:AD816} (and Example \ref{eg:ternary423}):
\begin{example}
\label{eg:GF3}
The code given in Example \ref{eg:AD816} under the $\mathfrak{S}$ map (pairing up coordinates $\{1,2\},\{3,4\},\{5,6\},\{7,8\}$) gives a linear code over $GF(3)$ generated by $\{0111,1012\}$.
\end{example}
We know that all the linear ternary codes are self-complementary, so the $1$-codes constructed from linear ternary codes of distance $3$ can directly used to construct single-error-correcting AD codes \cite{gf3classical}. Since in general we search for self-complementary ternary codes ${\mathcal C}'$ with largest possible size of ${\mathcal C}=\mathfrak{S}({\mathcal C}')$, those AD codes obtained from linear ternary codes of distance $3$ are sub-optimal.

We now show that the AD codes given by the Constantin-Rao construction are actually a special case of the $GF(3)$ construction.


\begin{theorem}
\label{th:ternary}
For $n$ even, the Varshamov-Tenengol'ts code ${\mathcal V}_0$, and the Constantin-Rao code ${\mathcal C}_0$ of largest cardinality has ternary form. 
\end{theorem}
\textbf{Proof}
We only need to prove that there exists a choice of pairing, such that for any codeword 
$v\in \mathcal{V}_0\ (\mathcal{C}_0)$, if $v$ restricts on one chosen pair $\alpha$ is $00$, then there exists another codeword 
$v'\in \mathcal{V}_0\ (\mathcal{C}_0)$ such 
that $v'=v|_{\tilde{\alpha}}$  and $v'|_{\alpha}=11$. Here $\tilde{\alpha}$ 
denotes all the other coordinates apart from $\alpha$. 

For the Varshamov-Tenengol'ts code $\mathcal{V}_0$ of even length $n$,
choose the pairing $\{i,n-i+1\}_{i=1}^{n/2}$, then the above condition
is satisfied. This is because $i+n-i+1=n+1\ mod\ n+1=0$.

For the Constantin-Rao code $\mathcal{C}_0$ of largest cardinality,
which is given by the group
$G=\bigoplus_{r}\bigoplus_{i=1}^{n_{r}}\mathbb{Z}_{p_r}$, note $n$ is
even, so $n+1$ is odd. Therefore all $p_r$ are odd for $p_r|n+1$,
where $n+1=\Pi_{p_r|n+1}p_r^{n_{r}}$. Write any group element as
$(s_{11},...,s_{1{n_1}},s_{21},...,s_{2{n_2}}... )$. Then we can pair
it with
$(p_1-s_{11},...,p_1-s_{1{n_1}},p_2-s_{21},...,p_2-s_{2{n_2}}... )$,
$mod\ (p_1,...,p_1,p_2,...,p_2,...)  $, where
$s_{rj_r}\in\{0,...,p_r-1\}$ and $j_r=1,...,n_r$.  $\square$

From both Fact \ref{fact:even} and Theorem \ref{th:ternary} we learn that for even block length, the Constantin-Rao code ${\mathcal C}_0$ of maximal cardinality is both self-complementary and has ternary form. Therefore, the AD codes given by the Constantin-Rao construction is actually a special case of the $GF(3)$ construction. 

\subsubsection{Odd block length}

For $n$ odd, we need to generalize the $GF(3)$ construction. As already discussed in Sec. \ref{oddonecode}, for $n=2m+1$, we design codes correcting a single error of the channel ${\mathcal Z}\times{\mathcal T}^{m}$. And we call these codes `generalized ternary.'

We need to examine under which condition the image of a generalized ternary code under $\mathfrak{S}$ is self-complementary.

\begin{definition}
A generalized ternary code ${\mathcal C}'$ of length $2m+1$ is self-complementary if for any $\mathbf{c}'\in {\mathcal C}'$, $\bar{\mathbf{c}}'\in {\mathcal C}'$. Here $\bar{{c}}'_1=1\oplus {c}'_1$, $\bar{{c}}'_i=3\ominus {c}'_i\ \mod 3$, for $i=2,\ldots,m+1$.
\end{definition}
\begin{example}
The generalized ternary code ${\mathcal C}'=\{000,100,011,122\}$ is self-complementary, because $\overline{000}=100$ and $\overline{011}=122$.
\end{example}
The properties of $\mathfrak{S}$ give the following:
\begin{fact}
\label{fact:getecomp}
If a generalized ternary code ${\mathcal C}'$ of length $m+1$ is self-complementary, then its binary image under the map ${\mathcal C}=\mathfrak{S}^m({\mathcal C}')$ is self-complementary. On the other hand, if a binary code ${\mathcal C}$ of length $2m+1$ has generalized ternary form and is self-complementary, then its image $\tilde{\mathfrak{S}}^{2m}({\mathcal C})$ is self-complementary.
\end{fact}

We now show that the AD codes given by the Constantin-Rao construction are actually a special case of the generalized ternary construction. 

\begin{definition}
A binary code ${\mathcal C}$ of odd length $n=2m+1$ has {\em generalized ternary form} if $\mathfrak{S}^{m}(\tilde{\mathfrak{S}}^m({\mathcal C}))={\mathcal C}$.
\end{definition}

Based on this definition, if a binary code ${\mathcal C}$ of odd length $2m+1$ has generalized ternary form, then it can be constructed from some codes correcting a single error of the channel ${\mathcal Z}\times{\mathcal T}^m$ via the ternarization map. The following theorem then shows that certain Varshamov-Tenengol'ts-Constantin-Rao codes are a special case of asymmetric codes constructed from single-error-correcting codes for the channel ${\mathcal Z}\times{\mathcal T}^m$ \cite{gf3classical}.

\begin{theorem}
\label{th:geternary}
For $n$ odd, the Varshamov-Tenengol'ts code ${\mathcal V}_g$ has generalized ternary form.
\end{theorem}
\textbf{Proof}
We only need to prove that there exists a choice of pairing which leaves a single coordinate as a bit, such that for any codeword $v\in \mathcal{V}_g$, if $v$ contains the paired bits $00$, then there exist another codeword $v'\in \mathcal{V}_g$ which is identical except that the $00$ pair is replaced by $11$, and vice versa.

For the Varshamov-Tenengol'ts code $\mathcal{V}_g$ of odd length, choose the pairing $\{i,n-i+1\}_{i=1}^{(n-1)/2}$, leave the coordinate $(n+1)/2$ as a bit,
 then the above pairing condition is satisfied. This is because $i+(n-i)+1=(n+1)\!\mod (n+1)=0$. $\square$

Now recall Fact \ref{fact:4k3}, which states that for block length $n=4k+3$, ${\mathcal V}_{\frac{n+1}{4}}$ is self-complementary. We further show the following:
\begin{fact}
For $n=4k+3$, ${\mathcal V}_{\frac{n+1}{4}}$ is of generalized ternary form.
\end{fact}
To see this, do the pairing $\{i,n-i+1\}_{i=1}^{(n-1)/2}$. Here we leave the coordinate $(n+1)/2$ unpaired so it is unchanged under the map $\tilde{\mathfrak{S}}^m$.

For length $4k+1$, recall Fact \ref{fact:4k1} that the shortened Varshamov-Tenengol'ts code ${\mathcal V}'_{\frac{n+2-r}{2}}$ obtained by deleting any `odd' coordinate $r$ from Varshamov-Tenengol'ts code ${\mathcal V}_{\frac{n+2-r}{2}}$ of length $n+1$ is self-complementary. We further show the following:
\begin{fact}
For $n=4k+1$, the shortened Varshamov-Tenengol'ts code ${\mathcal V}'_{\frac{n+2-r}{2}}$ obtained by deleting any `odd' coordinate $r$ from Varshamov-Tenengol'ts code ${\mathcal V}_{\frac{n+2-r}{2}}$ of length $n+1$ has generalized ternary form.
\end{fact}
To see this, for the shortened Varshamov-Tenengol'ts code given by
\begin{equation}
\sum_{i=1,i\neq r}^{n+2} ix_i=\frac{n+2-r}{2} \mod\ n+2,
\end{equation}
do the pairing $\{i,n-i+2\}_{i=1}^{n/2}$. Here we leave the coordinate $n-r+2$ unpaired so it is unchanged under the map $\tilde{\mathfrak{S}}^m$.

\section{Summary of new constructions for amplitude damping codes \label{summary}}
\label{sec:summary}

For short block length we summarize the results of single-error-correcting AD codes obtained from the $GF(3)$ construction in Table \ref{table:ADcodes}, and compare them with AD codes obtained from other constructions.

\begin{table}[htbf]
\caption{Codes:
This table compares the various constructions for amplitude damping 
codes, giving the best known codes created by various constructions. 
The first column gives the number of qubits. The
second column gives additive codes. The third column uses the construction
given in Gottesman \cite{Daniel}. The third column gives 
codes created by the complementary construction of Lang and Shor
\cite{Peter2}.  The fourth column (${\mathcal C}_g$) gives Constantin-Rao codes.  The
fifth column gives codes constructed using Theorem~\ref{th:ADCSS} and
computer search.}
\begin{equation}\nonumber
\begin{array}{c c c c c c}
\\
\hline
n & GF(4) & \cite{Daniel} &  \cite{Peter2} & {\mathcal C}_g & GF(3)\\
\hline
4 & 1 & 1 &2 & 2 & 2\\
5 & 2 & 2 &2 & 2 & 2\\
6 & 2 & 4 &5 & 5 & 5\\
7 & 2 & 8 &8 & 8 & 8\\
8 & 8 & 8 &12 & 16 & 16\\
9 & 8 & 16 &18 & 23 & 24 \\
10 & 16 & 32 &41 & 47 & 49\\
11 & 32 & 64 & 78 & 86 & 89 \\
12 & 64 & 128 &146 & 158 & 168\\
13 & 128 & 256 & 273 & 274 & 291 \\
14 & 256 & 512 & 515 & 548 & 572\\
15 & 512 & 1024 & 931 & 1024 & * \\
16 & 1024 & 1024 & 1716 & 1928 & * \\
\hline
\end{array}
\end{equation}

\label{table:ADcodes}
\end{table}

Note the $\lfloor\lfloor 12,168,1\rfloor\rfloor$ code in Table
\ref{table:ADcodes} is cyclic, which can be obtained by the classical
$1$-code $\lfloor 12,336,1\rfloor$ given in \cite{gf3classical}. The
$\lfloor\lfloor 10,49,1\rfloor\rfloor$ code is `almost cyclic', from
which (deleting $4$ classical codewords then add another $2$) we can
obtain a cyclic code $\lfloor\lfloor 10,47,1\rfloor\rfloor$, with
classical codewords
\begin{equation}
00000\ 11111\ 22222\ 21100\ 20111
\end{equation}
and their cyclic shift, plus all the complements. There is another cyclic code  $((10,47))$, with classical codewords
\begin{equation}
00000\ 11111\ 22222\ 21100\ 21011
\end{equation}
and their cyclic shift, plus all the complements. 

Table \ref{table:ADcodes} shows that the Constantin-Rao construction ${\mathcal C}_g$ outperforms other constructions apart from the (generalized) $GF(3)$ construction. This is reasonable since we know that the Constantin-Rao construction is actually a special case of the (generalized) $GF(3)$ construction. For all lengths up to $14$, the (generalized) $GF(3)$ construction indeed gives AD codes of best parameters. Lengths $>14$ are out of reach of the current computational power we have. As we know that the Constantin-Rao construction outperform the CSS construction for all lengths except $n=2^r-1$, where the binary Hamming codes are `good', it is very much desired to know whether the (generalized) $GF(3)$ construction can give us something outperforms the CSS construction for the length $n=2^r-1$. From \cite{gf3classical} we know this is possible for classical $1$-codes, but it remains a mystery for the quantum case, which we leave for future investigation. 

Finally, numerical search also found a $\lfloor\lfloor 9,26,1\rfloor\rfloor$ single-error-correcting AD code (exhaustively found to be optimal among all the self-complementary codes), which cannot be  obtained from any of the above constructions. Also we have found, via random search, a $\lfloor\lfloor 10,51,1\rfloor\rfloor$ code, which also cannot be obtained from any of the above constructions.

\section*{Acknowledgements} GS and JAS received support from the
DARPA QUEST program under contract no. HR0011-09-C-0047.


\end{document}